\documentclass[11pt]{article}  %>>> use for US letter paper
\usepackage{booktabs}
\usepackage{cite}
\usepackage[letterpaper,margin=1in]{geometry}
\usepackage{authblk}
\usepackage{amsmath,amssymb}
\usepackage{graphicx}
\usepackage{hyperref}
\usepackage{amsmath,amssymb,amsfonts}
\usepackage{algorithmic}
\usepackage{graphicx}
\usepackage{textcomp}
\usepackage{subfig}
\usepackage{textcomp}
\usepackage{stfloats}
\usepackage{url}
\usepackage{verbatim}
\usepackage{cite}
\usepackage{amsmath,adjustbox}
\usepackage[table,xcdraw]{xcolor}
\usepackage{graphicx,xcolor} 
\usepackage{booktabs}
\usepackage[table,xcdraw]{xcolor}
\usepackage{caption}
\usepackage{hyperref}
\usepackage{subcaption}

\title{Constrained and Regularized Quantitative Ultrasound Parameter Estimation using ADMM}

\author[1]{Ali K. Z. Tehrani}
\author[1]{Hassan Rivaz}
\author[2]{Ivan M. Rosado-Mendez}

\affil[1]{Department of Electrical and Computer Engineering, Concordia University, Canada}
\affil[2]{Departments of Medical Physics and Radiology, University of Wisconsin, United States}

\begin{document} 
	\maketitle
	
	\begin{abstract}
Regularized estimation of quantitative ultrasound (QUS) parameters, such as attenuation and backscatter coefficients, has gained research, clinical, and commercial interests. Recently, the alternating direction method of multipliers (ADMM) has been applied successfully to estimate these parameters by utilizing L2 and L1 norms for attenuation and backscatter coefficient regularization, respectively. While this method improves upon previous approaches, it does not fully leverage the prior knowledge of minimum physically feasible parameter values, sometimes yielding values outside the realistic range. This work addresses this limitation by incorporating minimum QUS parameter values as constraints to enhance ADMM estimation. The proposed method is validated using experimental phantom data. 
	\end{abstract}
	
	% Include a list of keywords after the abstract 
\begin{center}
\small\textbf{Keywords:} Quantitative ultrasound; Tissue characterization; ADMM; Constrained optimization
\end{center}
\section{Introduction}	
Quantitative ultrasound (QUS) reveals the quantitative properties of the tissue microstructure, and has been extensively utilized for breast cancer diagnosis \cite{quiaoit2020quantitative}, fatty liver disease classification \cite{nguyen2019reference}, prostate cancer monitoring \cite{rohrbach2018high}, among other applications. Spectral-based QUS methods employ radiofrequency (RF) data backscattered from the tissue to estimate different parameters such as attenuation coefficient, backscattering coefficient (BSC), and effective scatterer diameter (ESD) \cite{insana1990characterising,oelze2016review,nam2011ultrasonic,oelze2002method,mamou2013backscatter,ghoshal2013state,lye2021vivo}. In this work, we focus on estimating the attenuation coefficient and BSC from the power spectra.     

Recent work has also explored deep learning (DL) for QUS to learn mappings from backscattered signals to tissue properties and labels. For example, DL-based classifiers have been used to discriminate scatterer density regimes directly from RF data, reducing the reliance on large patches and reference phantoms while improving robustness to different imaging settings~\cite{tehrani2021ultrasound}. In another work, a spectral network called SincNet, originally proposed for speech processing, was modified and adopted to ultrasound RF data to classify ultrasound microstructure \cite{tehranispeech}. DL-based methods have also been widely utilized in attenuation coefficient estimation \cite{marin2024deep,liu2025deep,timana2024deep}.
 Data-efficient training methods like zone-based schemes further reduce the amount of labeled data required for tissue characterization via QUS \cite{soylu2023data}. Despite this progress, optimization-based QUS methods are inherently explainable and are the focus of this paper.

Several approaches have been proposed to estimate the mentioned parameters. Reference phantom method (RPM) is a well-known method used to cancel the system-dependent effects \cite{yao1990backscatter}. In \cite{vajihi2018low}, dynamic programming was introduced to account for the dependency between neighboring samples, incorporating regularization to mitigate variance. Another method, Analytical Globally Regularized Backscatter Quantitative Ultrasound (ALGEBRA), estimates the parameters analytically. It optimizes a penalty function with L2-norm regularization terms \cite{jafarpisheh2020analytic}. Other cost-function–based approaches include Refs.~\cite{coila2017regularized,chahuara2020regularized,romero2020regularized}. More recently, the Alternating Direction Method of Multipliers (ADMM) has been suggested to overcome the limitations of solving the problem in ALGEBRA, such as relying on the L2 norm for both attenuation and BSC, assigning equal importance to all frequencies, and reduced performance in the presence of specular reflectors \cite{jafarpisheh2023physics}.
This work presents a modification to the ADMM solution to incorporate prior knowledge of the minimum value of the estimated parameters as a constraint in the optimization framework to reduce artifacts when the observation is unreliable. The demo code is publicly available at \url{http://code.sonography.ai/}.  

\section{Material and Methods}
\subsection{Problem formulation}
Based on RPM, the power spectra from RF signals obtained from the target of interest are divided by the power spectra of a reference phantom with known parameters to remove system-dependent effects, which yields \cite{yao1990backscatter,vajihi2018low,jafarpisheh2020analytic,jafarpisheh2023physics}:
\begin{equation}
	\label{eq1}
	\frac{S_{t}(x,z,f)}{S_{r}(x,z,f)}=\frac{\sigma_t(x,z,f)A_t(x,z,f)}{\sigma_r(x,z,f)A_r(x,z,f)}
\end{equation}
where $x$, $z$, and $f$ denote the lateral position, depth, and frequency, respectively, and the subscripts $t$ and $r$ represent the target and reference phantom. The system-independent power spectra are modeled as a multiplication of the BSC ($\sigma$) and total attenuation ($A$), which can be parametrized as \cite{yao1990backscatter,vajihi2018low,jafarpisheh2020analytic,jafarpisheh2023physics}: 
\begin{equation}
	\label{eq:alpha_sigma}
	\begin{gathered}
		\sigma_t (z,f) = b_t(z)f^{n_t(z)}\\
		A_t(z,f) = exp(-4\overline{\alpha_t}fz)
	\end{gathered}
\end{equation}
by taking the natural logarithm and applying (\ref{eq:alpha_sigma}) into (\ref{eq1}) we can obtain:
\begin{equation}
	\label{eq_qus1}
	\begin{gathered}
		X(f, z) = Ln(\frac{S_{t}(x,z,f)}{S_{r}(x,z,f)}) = b + n \ln f - 4 \alpha f z \\
		\ln \frac{b_t}{b_r} \equiv b, \quad n_t - n_r \equiv n, \quad \overline{\alpha_t} - \overline{\alpha_r} \equiv \alpha
	\end{gathered}
\end{equation}
where $b_t$ is the backscattering coefficient of the target, $\overline{\alpha_t}$ is the path-averaged attenuation coefficient (assumed linear in frequency), and $n_t$ characterizes the frequency dependence of backscattering.  The goal is to estimate $b$, $n$, and $\alpha$ from the observed $X(f,z)$. 
\subsection{ADMM method}
Jafarpisheh \textit{et al.} proposed using ADMM optimizer to estimate the QUS parameters. They used L2 norm regularization for $\alpha$ since it has gradual variations and employed L1 norm regularization for $b_t$ and $n_t$ since they can have abrupt changes \cite{jafarpisheh2023physics}. They also utilized adaptive weights for different frequencies since the importance of different frequencies may change with depth and noise level.

Explaining the details of how the ADMM solution has been derived is out of scope of this work but the final results are given here. The algorithm tries to minimize the following constrained cost function:
\begin{equation}
	\label{eq:const}
	\begin{aligned}
		C &= \frac{1}{2} \| Hx - t \|_2^2 + \lambda_1 \| s_1 \|_2^2 + \lambda_2 \| s_2 \|_1 \\
		\text{s.t.} \quad &K_1 x_1 - s_1 = 0, \quad K_2 x_2 - s_2 = 0.
	\end{aligned}
\end{equation}
where $x$ is the vector containing $b$, $n$, $\alpha$ of all samples in a line, $x = \left\{\alpha_1,\alpha_2,...,\alpha_{N_R},b_1,b_2,...,b_{N_R}, n_1,n_2,...,n_{N_R} \right\}^T$, and $N_R$ denotes the number of samples in a line. $H$ is a known transformation matrix given in Eq \ref{eq:h} (refer to \cite{jafarpisheh2023physics} for more details of how it was derived), and $t$ is the observed power spectra sample.
\begin{equation}
\label{eq:h}
H  =
\begin{bmatrix}
\operatorname{diag}\!\Big(\sum\nolimits_f (4 z f)^2\Big) &
\operatorname{diag}\!\Big(\sum\nolimits_f (4 z f)\Big) &
\operatorname{diag}\!\Big(\sum\nolimits_f (4 z f)\ln f\Big) \\
\operatorname{diag}\!\Big(\sum\nolimits_f (4 z f)\Big) &
\operatorname{diag}\!\Big(\sum\nolimits_f 1\Big) &
\operatorname{diag}\!\Big(\sum\nolimits_f \ln f\Big) \\
\operatorname{diag}\!\Big(\sum\nolimits_f (4 z f)\ln f\Big) &
\operatorname{diag}\!\Big(\sum\nolimits_f \ln f\Big) &
\operatorname{diag}\!\Big(\sum\nolimits_f (\ln f)^2\Big)
\end{bmatrix},
\end{equation}
The $x_1$ is part of $x$ containing only $\alpha$ values ($x_1 = \left\{\alpha_1,\alpha_2,...,\alpha_{N_R}\right\}^T$), $x_2$ is the part containing $b$ and $n$ values ($x_2=\left\{b_1,b_2,...,b_{N_R}, n_1,n_2,...,n_{N_R} \right\}^T$); and $K_1$, and $K_2$ denote the regularization weight matrices. By introducing $x_1$ and $K_1$, the $\alpha$ is separated from $b$ and $n$ to impose L2 norm regularization due to a smoother variation compared to $b$ and $n$. By adding the latent variable ($s=\left\{s_1,s_2\right\}$),  (\ref{eq:const}) can be converted to an unconstrained cost function, and ADMM update rules can be derived as:
\begin{equation}
	\begin{aligned}
		x^{k+1} &:= \left( H^T H + \rho K^T K \right)^{-1} H^T t + \rho K^T \left( s^k - y^k \right) \\
		s_1^{k+1} &:= \left( K_1 x_1^{k+1} + y_1^k \right) / (\rho + \lambda_1) \\
		s_2^{k+1} &:= S_{\lambda_2 / \rho} \left( K_2 x_2^{k+1} + y_2^k \right) \\
		y^{k+1} &:= y^k + K x^{k+1} - s^{k+1}
	\end{aligned}
\end{equation}
where $S_{\frac{\lambda}{\rho}} = \text{sgn}( \cdot ) \max \left\{ \left| \cdot \right| - \frac{\lambda}{\rho}, 0 \right\}
$ performs soft thresholding.  
\subsection{Proposed method}
The ADMM approach incorporates the regularization of L1 and L2 norms, but it does not utilize any constraint regarding the feasible range of the estimated variables. In this work, we incorporate the minimum physically feasible values as the constraints to provide prior knowledge to the cost function. The new constrained optimization cost function can be written as \cite{buzzard2018plug}:
\begin{equation}
	\label{eq:const2}
	\begin{aligned}
		C &= \frac{1}{2} \| Hx - t \|_2^2 + \lambda_1 \| s_1 \|_2^2 + \lambda_2 \| s_2 \|_1 \\
		\text{s.t.} \quad &K_1 x_1 - s_1 = 0, \quad K_2 x_2 - s_2 = 0.\\
		\quad\quad&\Phi x-v= \beta
	\end{aligned}
\end{equation}

The last phrase is added to represent the minimum value constraint,\\ where $\beta = \left[\alpha_{m_1}, \dots, \alpha_{m_{N_R}}, b_{m_1}, \dots, b_{m_{N_R}}, n_{m_1}, \dots, n_{m_{N_R}} \right]^T $ is the vector containing the minimum feasible values for each parameter, $v$ denotes the new latent variable added to facilitate the incorporation of the constraint, and $\Phi$ is an identity matrix. We considered that the attenuation ($\overline{\alpha_t}$), BSC ($b_t$), and $n_t$ are greater than 0, 0.001 of the reference and 0, respectively. Therefore, the minimum values ($\alpha_m$, $b_m$, and $n_m$) can be obtained as: 
\begin{equation}
	\label{eq:min}
	b_m \equiv \ln(0.001) , \quad n_m\equiv-n_r , \quad \alpha_m \equiv- \overline{\alpha_r} 
\end{equation}
The constrained optimization in \ref{eq:const2} can be converted into an unconstrained optimization problem by adding a latent variable (called $q$ here), and the update rule can be written as \cite{buzzard2018plug}: 
\begin{figure}[tb]	
	\centering
	\includegraphics[width=0.37 \textwidth]{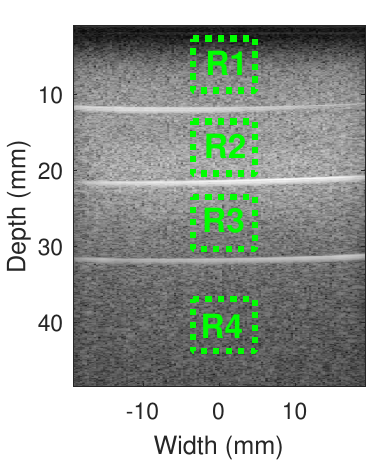}
	\caption{The B-mode image of the phantom. }
	\label{fig:specular_bmode}	
\end{figure}
\begin{equation}
	\begin{aligned}
		x^{k+1} &:= \left( H^T H + \rho K^T K \right)^{-1} H^T t \\
		&\quad + \rho K^T \left( s^k - y^k \right) + \gamma \phi^T (v^k - q^k) \\
		v^{k+1} &= \text{clip} \{ x^{k+1} + q, \beta \} \\
		q^{k+1} &= q^k + \left( x^{k+1} - v^{k+1} \right) \\
		s_1^{k+1} &= \frac{K_1 x_1^{k+1} + y_1}{\rho + \lambda_1} \\
		s_2^{k+1} &= s_2 \left( K_2 x_2 + y_2 \right) \\
		y &= y + K x^{k+1} - s
	\end{aligned}
\end{equation}
The clip function ensures that $v$ stays higher than the minimum physically feasible value ($\beta$), and the latent variable $q$ tries to make the estimated variables ($x$) as close as possible to $v$. We named this method Constrained-ADMM (C-ADMM), and $\gamma$ is a hyper-parameter that should be tuned for each application. 
\subsection{Datasets}

We employed the Verasonics Vantage 128 system (Verasonics, Kirkland, WA, USA) with an L11-5v transducer operating at 8 MHz to collect data from  the Gammex 410 SCG phantom
(Gammex-Sun Nuclear, Middleton, WI, USA, serial number
805546-4612). There are nylon filaments to produce specular reflection, and the transducer aperture was aligned with the long axis of the fibers. The phantom has $\overline{\alpha_t}=0.6035$ $dB$ $cm^{-1}$ $MHz^{-1}$, $b_t=2.996\times 10^{-6}$ $cm^{-1}$ $sr^{-1}$, and  
$n_t = 3.428$. The B-mode image of this phantom is depicted in Fig.  \ref{fig:specular_bmode}. The main interest is estimating the QUS parameters of the background in the presence of reflectors to investigate how the reflectors affect the estimation of the background QUS parameters.

%\subsubsection{Phantom With Inclusions}
%We also evaluated the algorithm's performance using a different region of the previous phantom having three cylindrical inclusions with +12 dB, +6 dB, and -6 dB backscattering coefficients with respect to the background. Other parameters are the same as the phantom with specular reflectors.

\subsection{Quantitative metrics}
We report the bias error and variance error of the estimated parameters, which can be defined as:
\begin{equation}
	bias = E\left\{|\theta -\theta_{gt} |\right\}, \quad Variance = E\left\{ (\theta-\theta_{gt})^2\right\}
\end{equation}
where $\theta$ is the estimated parameter, and $E(.)$ denotes averaging within the ROI. We report the bias and variance of BSC in dB scale. The results are reported for the four regions of size $5\times 6 mm$ highlighted in Fig. \ref{fig:specular_bmode}.

\begin{figure}[t]	
	\centering
	\includegraphics[width=0.6\textwidth]{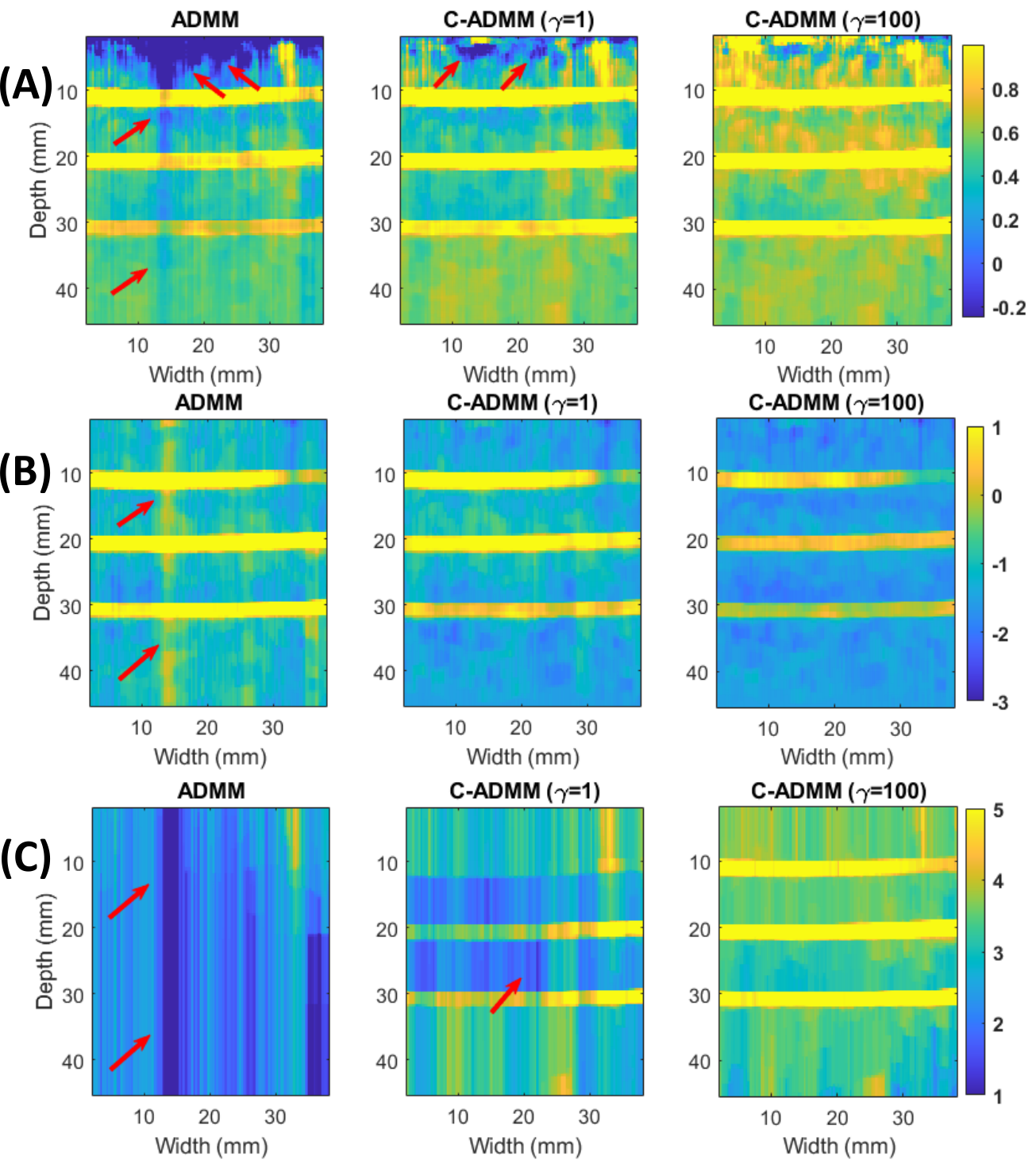}
	\caption{QUS parameter estimation results. $\alpha_t:$ Attenuation Coefficient (A), $b_t:$ Backscattering Coefficient in dB (B), and $n_t:$ frequency dependency (C). The marked area with red arrows illustrates the region where there are artifacts.}
	\label{fig:specular_res}	
\end{figure}
%\begin{figure}[ht]	
%	\centering
%	\includegraphics[width=0.25\textwidth]{att_bias.png}
%	\caption{Attenuation Coefficient ($\alpha_t$) bias error.}
%	\label{fig:bias_at}	
%\end{figure}
%
%\begin{figure}[ht]	
%	\centering
%	\includegraphics[width=0.25\textwidth]{bsc_bias.png}
%	\caption{Backscattering Coefficient ($b_t$) bias error.}
%	\label{fig:bias_b}	
%\end{figure}
%\begin{figure}[ht]	
%	\centering
%	\includegraphics[width=0.25\textwidth]{n_bias.png}
%	\caption{Frequency dependency power ($n_t$) bias error. }
%	\label{fig:bias_n}	
%\end{figure}

\begin{figure*}[t]
	\centering
	% --- (b) Bias α ---
	\begin{minipage}[t]{0.325\textwidth}
		\centering
		\includegraphics[width=\textwidth]{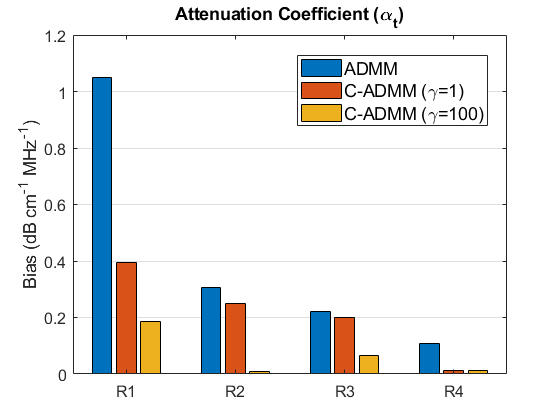}
		\vspace{2mm}
		{\small\textbf{(a)}}
	\end{minipage}
	\hfill
	% --- (c) Bias b ---
	\begin{minipage}[t]{0.325\textwidth}
		\centering
		\includegraphics[width=\textwidth]{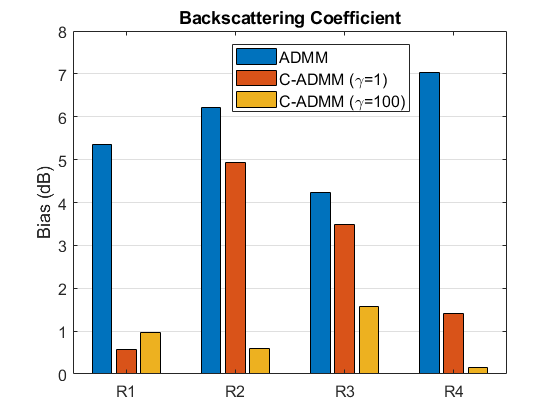}
		\vspace{2mm}
		{\small\textbf{(b)}}
	\end{minipage}
	\hfill
	% --- (d) Bias n ---
	\begin{minipage}[t]{0.325\textwidth}
		\centering
		\includegraphics[width=\textwidth]{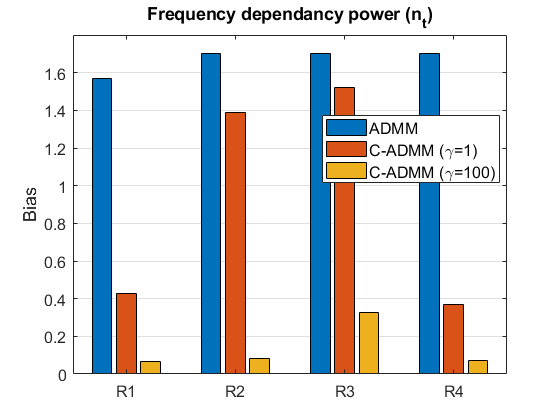}
		\vspace{2mm}
		{\small\textbf{(c)} }
	\end{minipage}
	
	\caption{ 
		 Bias error analysis of the quantitative ultrasound (QUS) parameters.  
		(a) Attenuation coefficient bias ($\alpha_t$),  
		(b) backscattering coefficient bias ($b_t$), and  
		(c) frequency–dependence power bias ($n_t$).
	}
	\label{fig:combined}
\end{figure*}

\section{Results}

Figure \ref{fig:specular_res} shows the estimated attenuation coefficient map (A), BSC in dB (B), and $n_t$ (C) using ADMM, the proposed  C-ADMM with $\gamma=1$, and $\gamma=100$. The red arrows illustrate the region where there is a large error. ADMM has a high error in depths lower than 10 mm for the attenuation coefficient, giving negative values that are not physically feasible.  Furthermore, an incorrect band is visible in all three estimated parameters by ADMM. The proposed C-ADMM with $\gamma=100$ does not estimate negative values for the attenuation coefficient, and the incorrect band has been removed. It should also be noted that ADMM fails to correctly calculate $n_t$ (C) in all regions. However, C-ADMM obtains a more reliable estimate of this parameter. Comparing the C-ADMM with $\gamma=1$ and $\gamma=100$,  C-ADMM with $\gamma=100$ outperforms C-ADMM with $\gamma=1$ (artifacts are marked in C-ADMM with $\gamma=1$ by red arrows).    
\begin{table*}[t]
	\centering
	\caption{Variance error of the estimated parameters}
	\resizebox{0.95\textwidth}{!}{%
		\begin{tabular}{@{}c|cccc|cccc|cccc@{}}
			\toprule
			& \multicolumn{4}{c|}{$\alpha_t$} & \multicolumn{4}{c|}{$b_t$} & \multicolumn{4}{c}{$n_t$} \\ \cmidrule(l){2-13} 
			Method & R1 & R2 & R3 & R4 & R1 & R2 & R3 & R4 & R1 & R2 & R3 & R4 \\ \midrule
			ADMM & 0.343 & 0.018 & 0.007 & 0.002 & 10.635 & 13.770 & 7.970 & 11.098 & 0.408 & 0.358 & 0.358 & 0.358 \\
			C-ADMM ($\gamma=1$) & \textbf{0.074} & \textbf{0.009} & \textbf{0.001} & 0.002 & 4.918 & 7.067 & 4.080 & 2.369 & 0.071 & 0.062 & \textbf{0.028} & 0.118 \\
			C-ADMM ($\gamma=100$) & 0.076 & 0.013 & 0.004 & \textbf{0.001} & \textbf{1.700} & \textbf{1.461} & \textbf{1.061} & \textbf{0.663} & \textbf{0.016} & \textbf{0.047} & 0.037 & \textbf{0.050} \\ \bottomrule
		\end{tabular}%
	}
	\label{tab:my-table}
\end{table*}

The bias errors for different regions specified in Fig. \ref{fig:specular_bmode} are presented in Fig. \ref{fig:combined} (a), \ref{fig:combined} (b) and \ref{fig:combined} (c) for the attenuation coefficient, BSC, and $n_t$, respectively. As shown in Fig. \ref{fig:combined} (a), region R1 exhibits a substantial bias error in attenuation estimation, particularly with the ADMM method. However, C-ADMM significantly reduces this bias error, providing estimates that are much closer to the ground truth in this region. Additionally, for BSC and $n_t$, ADMM consistently shows high bias error across all regions. When comparing the hyper-parameter $\gamma$ values, C-ADMM with 
$\gamma=100$ achieves considerably lower bias error than with $\gamma=1$, indicating that $\gamma=100$ is a more optimal choice.

The variance of the estimated parameters is reported for different regions in Table \ref{tab:my-table}. We can see a high variance in the attenuation coefficient estimated by ADMM in R1 (0.343), while C-ADMM has a considerably lower variance in this region (0.074 and 0.076). Furthermore, C-ADMM exhibits much lower variance for both BSC ($b_t$) and $n_t$ compared to ADMM, which showcases it as a more reliable method than ADMM.
% Below is an example of how to insert images. Delete the ``\vspace'' line,
% uncomment the preceding line ``\centerline...'' and replace ``imageX.ps''
% with a suitable PostScript file name.
% -------------------------------------------------------------------------

\section{Discussion}
In Jafarpisheh \textit{et al.} \cite{jafarpisheh2023physics}, the presented results of ADMM are averaged across multiple frames, which reduces the error. Herein, we did not perform any averaging across different frames to better demonstrate the effectiveness of the proposed regularization. Utilizing multiple frames further improves the accuracy of the estimated parameters. In addition, we employed a smaller patch size, which reduces averaging and better highlight the estimation error compared to \cite{jafarpisheh2023physics}.   

The main goal of imaging the phantom with nylon reflectors was to assess how strong specular reflectors influence the estimation of background QUS parameters, which is clinically important because many tissues contain highly reflective structures (e.g., calcifications, vessel walls, or interfaces) that can bias quantitative measurements of surrounding tissue if not properly accounted for. Our focus was therefore on recovering the attenuation and backscatter properties of the background in the presence of these reflectors, rather than characterizing the reflectors. Extending the proposed framework to explicitly model or estimate QUS parameters for highly specular structures remains an interesting direction for future work.

Computation-wise, the proposed C-ADMM only adds two operations in the ADMM update loop, which does not substantially increase the computation complexity compared to ADMM method. In our experiments, ADMM required 0.504 sec on average per line on an i7 CPU (11700F), whereas C-ADMM required 0.515 sec.   
%In this paper, we utilized the minimum feasible value of the estimated parameters as prior knowledge. The results are expected to improve when the optimizer fails to estimate physically feasible values due to noise and artifacts. We utilized the values given in \ref{eq:min}, which are general values for soft tissue. Depending on the application, the minimum value can be altered to provide more accurate prior knowledge to the optimizer.   

%\begin{figure}[htb]	
%	\centering
%	\includegraphics[width=0.45\textwidth]{gammex.pdf}
% \caption{Gammex phantom results. $\alpha_t:$ Attenuation Coefficient (A), $b_t:$ Backscattering Coefficient in dB (B), and $n_t:$ frequency dependency (C). }
%	\label{fig:gammex_res}	
%\end{figure}

\section{Conclusion}

This work introduced C-ADMM, a modification of ADMM that incorporates minimum value constraints to integrate prior knowledge into the optimization framework. The proposed approach demonstrated improved parameter estimation, particularly in regions where ADMM struggled to yield physically feasible values due to noise and artifacts.
\section*{Acknowledgment}
Funded by Government of Canada’s New Frontiers in Research Fund (NFRF), [NFRFE-2022-00295] and Natural Sciences and Engineering Research Council of Canada (NSERC).

% References should be produced using the bibtex program from suitable
% BiBTeX files (here: strings, refs, manuals). The IEEEbib.bst bibliography
% style file from IEEE produces unsorted bibliography list.
% ------------------------------------------------------------------------- 

%\bibliographystyle{spiebib} % makes bibtex use spiebib.bst
\bibliographystyle{IEEEbib}
\bibliography{refs} % bibliography data in report.bib	
\end{document}